\documentclass{kapproc} 
\usepackage{graphicx}
\kluwerbib

\begin{document}

\articletitle{Astrometry of circumstellar masers}

\author{H.J.~van Langevelde}
\affil{Joint Institute for VLBI in Europe\\
Postbus 2, 7990 AA Dwingeloo, the Netherlands} \email{langevelde@jive.nl}

\author{W.H.T. Vlemmings}
\affil{Sterrewacht Leiden\\
Niels Bohr weg 2, 2300 RA Leiden, The Netherlands} \email{vlemming@strw.leidenuniv.nl}

\begin{abstract}
The circumstellar masers around evolved stars offer an interesting
possibility to measure stellar parameters through VLBI astrometry. In
this paper the application of this technique is discussed, including
the accuracy and the uncertainties of the method. The
different maser species (OH, H$_2$O, SiO) have slightly different
characteristics and applications. This paper does not concern
astrometry of maser spots to study the kinematics of the envelope, but
concentrates on attempting to measure the motion of the underlying star.
\end{abstract}


\section*{Introduction}

There are a number of scientific justifications for astrometry of stars
with circumstellar masers. Historically, the registration of radio
single dish and infrared data was so uncertain that proper
identification of maser bearing stars was a problem. Most of these
problems have been overcome; the accuracy of connected element
interferometers is sufficient to make these identifications
unambiguously. However, there is still a lot of uncertainty when
comparing VLBI data from different experiments. Notably, when maser
distributions from different transitions are overlayed, a common origin
is usually assumed.  Astrometry can be done to do such comparisons
properly.

However, the most tantalizing application of astrometry of
circumstellar masers is the accurate determination of the motion of the
star on the sky. Proper motions of a star anywhere in the Galaxy (10
km/s at 8 kpc corresponds to $\mu = 0.25$ mas/year) can in principle be
obtained with VLBI, as can the parallax for such distances ($\pi=0.125$
mas).  A direct distance estimate is important for studying the physics
of the mass-losing stars. For many Mira variables Hipparcos measured
the distance, but for the most enshrouded stars VLBI offers the best
approach. Both the proper motion and distance are inputs for studying
the stellar population and Galactic kinematics (Habing, these
proceedings).

In addition, detailed astrometric monitoring of evolved stars could
reveal information about the existence of binary components, or even
planets in these systems. Moreover, by monitoring its position new
facts can be learned about maser physics. Are spots persistent, what
(turbulent) motions can be measured, is the brightness a function of
stellar cycle?  Finally it is worth mentioning that the maser
astrometry could tie the observational reference frames, in particular
in the infrared where these stars are very bright.

Carrying out astrometric monitoring projects with spectral line VLBI is
extensive work that requires some special techniques and assumptions.
There are a number of technical advancements and initiatives that will
make improvements to this field which will make it much more accessible.

\section{Methods \& Techniques}

The most straightforward way of doing VLBI astrometry is to use ``phase
referencing'' to a bright and close calibrator in order to overcome the
spatial and temporal fluctuations of the ionosphere and\slash or
troposphere.  The calibrator is usually an extragalactic source with a
fixed and known position, which is sufficiently bright to calibrate the
atmospheric and instrumental effects within a fraction of the coherence
time.  Because the modern VLBI correlators have incorporated the best
geodetic models, phase referencing results obtained in this way can be
accurate to a few mas.  This is already sufficient for a number of
applications. In other cases, higher accuracy is required and some
adjustments are necessary to improve the model. In such cases one
probably needs to return to the "totals" in order to use
astrometric/geodetic software. To get accuracies below 1 mas is not
trivial; better models of the atmospheric behavior and structure of the
reference source are required. 

For maser astrometry an additional complication can be that one needs
to observe in mixed bandwidth mode, applying a narrow band for the
maser detection and simultaneous wide bands for the reference source.
Obviously the maser dictates the observing frequency, which rules out
some calibration schemes used in continuum and geodetic VLBI.

Ideally the masers should be bright and persistent. The brightness 
relates to the size of individual maser features, which sets an
upper limit to the resolution one can obtain. A fundamental issue
remains to link the motion of the maser spot to the stellar
properties. One way forward is to assume that the masers are on a
linear path with respect to the star (e.g.\ a constant outflow). In
this case the parallax of individual spots equals the parallax of the
star, and the average motion may be assumed to be the stellar motion.
In other cases one may be able to deduce the position of the star from
the distribution of the masers, for example when they form a ring. If
we can determine the position of the star, one can in principle measure
the motion and the parallax.  Note that only when one can consistently
determine the stellar position with (much) better than 1 AU accuracy, a
useful parallax can be determined. In some cases there is evidence that
special maser spots exists that correspond to the stellar continuum
amplified by the maser shell. Such spots are then tied very accurately
to the stellar position. In other cases one has to worry about the
systematic and turbulent motions in the maser shell. In any case, there
may be additional motions involved, for example when masers occur in
binary stars.

\section{OH Masers}

The OH masers usually originate in large shells at 1000 AU from the
mass-losing stars. As the outflow is constant at that point, the maser
radiates most effectively radially, from the front and back cones of
the shell. Three 18 cm transitions can give reasonably bright maser
emission. However, the VLBI spots are usually resolved at 20 mas,
sometimes degraded further by angular broadening in the interstellar
medium, notably in the direction of the Galactic centre (Van Langevelde
et al.\ 1992).

Although the shells are large, the ``amplified stellar image'' paradigm
makes the OH maser interesting for astrometry. Radio continuum,
amplified by the maser results in a spot that is fixed on the stellar
position, bright and persistent. The observation that the most
blue-shifted spots in two OH transitions coincides is a very good
confirmation of this idea (Sivagnanam 1990). 

Recently, comparison of a proper motion study of U~Her with Hipparcos
agreed to 15 mas (van Langevelde et al.\ 2000), thus confirming that
the most blue-shifted and most persistent spot indeed follows the
stellar trajectory in this case. Therefore the maser motion can be used
to obtain the parallax and proper motion of the star. U~Her was
observed 10 times in 6 years on the NRAO VLBA which resulted in the first
VLBI parallax of a mass-losing star (Fig 1; Van Langevelde et al.\ 
2000).  The motion agrees within the error with the Hipparcos data,
from which no reliable parallax could be obtained.

If the amplified stellar image is a general feature in circumstellar OH
masers, it will allow astrometric measurements of a large number of
stars. However, the modest brightness of OH masers limits the
application of this technique to stars closer than 1 kpc.

\begin{figure}
  \centering
  \includegraphics[angle=-90,width=0.99\textwidth]{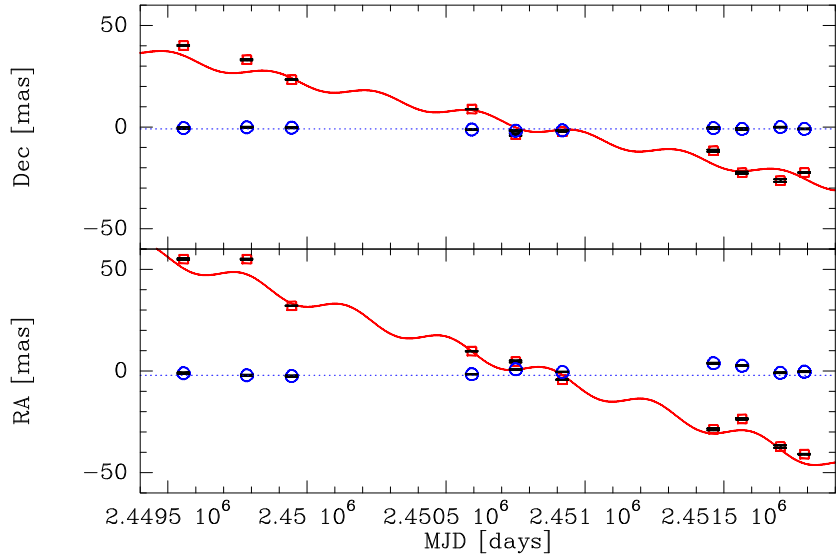}
  \caption{Proper motion of the most blue-shifted spot in the 1667 MHz
    maser of U~Her. A motion of $\mu_{\rm vlbi} = -15.57 \pm 0.56,
    -9.66 \pm 0.61$ mas/yr and a parallax of $\pi_{\rm vlbi} = 3.85
    \pm 1.14$ mas are fitted to the data. For comparison the position
    of a secondary reference source is also displayed.}
\end{figure}

\section{H$_2$O masers}

The 22 GHz masers are much brighter than the OH masers in circumstellar
shells. Therefore positional accuracies in the order of 50 $\mu$as are in
principle achievable. Of course at this high frequency the coherence
time is much shorter and phase referencing VLBI at 22 GHz is quite 
a bit more challenging.

Additional complications arise from the fact that the circumstellar
H$_2$O masers are rather messy. The structures are often ring-like on
scales of 100 AU, as expected from tangential amplification. However,
large departures have been observed, as well as considerable
variability. If the structures are persistent enough one can attempt to
track individual maser spots over a year and measure the parallax of
the star from these (e.g.\ Kuruyama et al., these proceedings).
If this approach works the water maser observations offer the
interesting possibility to measure parallaxes throughout a large
fraction of the Galaxy.

There have been claims that in water masers the amplified stellar image
can also occur (Marvel 1996, Colomer et al.\ 2000). To check this we
have observed U~Her at 22 GHz with MERLIN, which yields a comparable
resolution as our previous OH VLBI results. Using the same position
calibrators we find, somewhat surprisingly, that the brightest H$_2$O
maser coincides with the optical position and the OH image. Although
the MERLIN observations do not resolve individual maser spots
observable with VLBI, we take this as a clue that amplification of
background radiation may play a role in water masers too (Fig 2;
Vlemmings et al.\ 2002).

\begin{figure}
  \centering
  \includegraphics[width=0.99\textwidth]{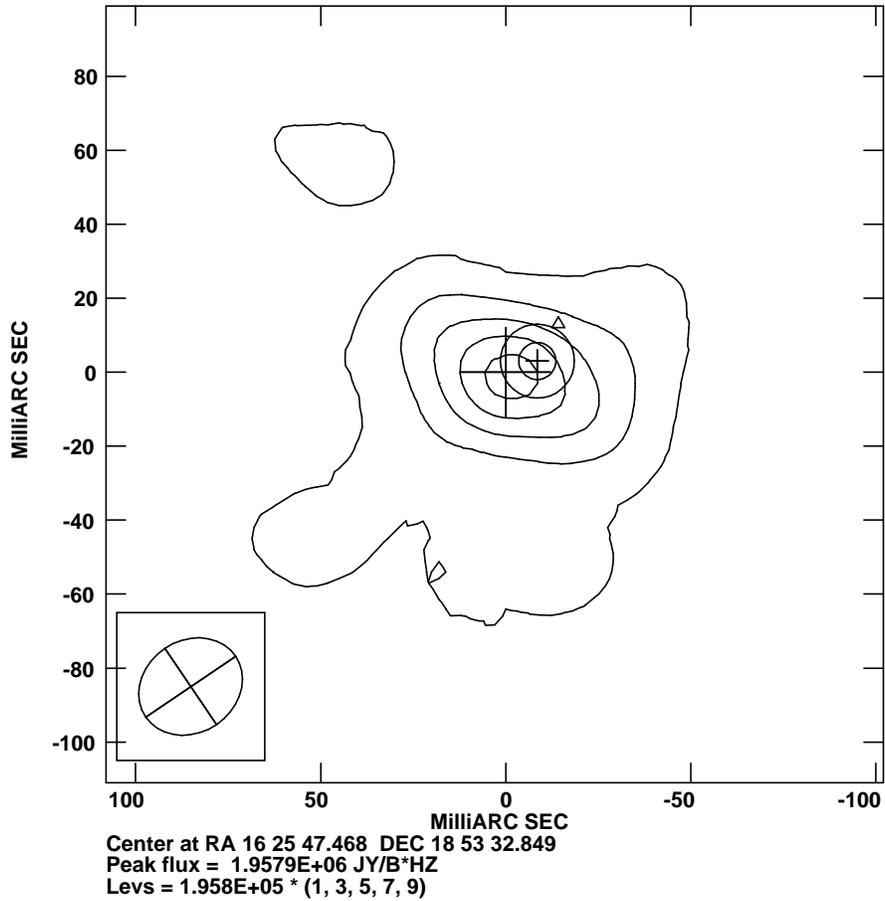}
  \caption{The location of the U~Her H$_2$O masers with respect to
    the stellar positions determined from the Hipparcos observations,
    as determined from phase referencing to an extragalactic
    calibrator. The star is denoted by two circles indicating the star
    itself and the stellar radio photo-sphere. The error bars on the
    maser feature are the errors due to the positional fitting. The
    errors on the stellar position are due to the link to the
    radio-reference frame and due to the errors in the proper motion
    used to transpose the optical position. The triangle denotes the
    stellar position when using the phase referencing results on the
    secondary calibrator. }
\end{figure}

\section{The use of SiO masers}

The SiO masers around evolved stars can be even brighter. At 43 GHz the
spots sizes allow 10$\mu$as accurate positions. At the same time the 
technical difficulties are considerable. The coherence time is only of
the order of a minute and in addition there are fewer suitable
extragalactic calibrators.

The masers themselves are located close to the star, within 10 AU. They
are bright and abundant. Monitoring of their structure indicates that the
masers are showing motions on the time-scale of the variability of the
star (Diamond \& Kemball 1999). Often these motions are not linear and it
will be a challenge to measure a parallax from the combined motions of
SiO masers.  As the masers are in a ring close to the star, it will be
possible to estimate the stellar position from the distribution of SiO
masers. The proper motion of the star can be measured this way, even if
the individual spots cannot be resolved. Proper motions of individual
stars can probably be measured throughout the Galaxy using the 43 GHz
transition.

A very suitable target for stellar astrometry by SiO masers are the
Galactic center stars. In this region there is an abundance of maser
stars close together (Sjouwerman, these proceedings). In addition,
SgrA* is a suitable calibrator; it is bright at short wavelengths and,
as it is is the location of a massive black hole, it is stationary. In
fact, an important results from circumstellar maser astrometry is the
tie between the IR and radio positions in the Galactic centre. By
locking the positions of the IR sources to the radio maser
counterparts, it can be shown that the motion of stars in the Galactic
centre is consistent with motion around a massive black hole located at
the position of SgrA* (Menten et al.\ 1997; Eckart et al.\ 2002).

Several groups have explored the possibility to start monitoring SiO
masers in the Galactic centre region with VLBI (Sjouwerman et al.\ 
1998; Deguchi, these proceedings; Imai, these proceedings; Reid,
private communication). The basic goal is to add two more components to
the measurement of kinematics in this dynamically interesting region
(Winnberg, these proceedings).

\section{Future}

There are many planned improvements of ordinary VLBI that will
facilitate more astrometry of circumstellar masers. Several modern, big
telescopes are under construction that will observe at high
frequencies, necessary for H$_2$O and SiO masers. Disk-based recording
systems will improve the reliability and productivity of VLBI, and will
be especially advantageous for time consuming astrometric monitoring.
In addition there are several indications that the atmospheric modeling
can be improved considerably. Other improvements of the geodetic models
are important too, for example accurate telescope positions.

However, the future for astrometry of masers clearly lies in
instrumental setups that overcome the temporal variability of the 
atmosphere at high frequencies. Some of this can be done with
``cluster--cluster'' VLBI, but the dedicated VERA project provides 
a far more ambitious approach that will soon produce interesting
results (Kobayashi \& Sasao, these proceedings).

A number of studies are required to establish new techniques for
astrometry of circumstellar masers. Careful analysis of errors and
consistency checks will be necessary to determine how to use the masers
optimally to determine the underlying stellar motions. But in potential
this method enables the measurement of proper motions and distances of
mass-losing stars throughout the Galaxy. This will progress the study
of stellar evolution as well as our knowledge of the structure of the
Galaxy.

\section{Acknowledgements}

The National Radio Astronomy Observatory is a facility of the National
Science Foundation operated under cooperative agreement by Associated
Universities, Inc.  MERLIN is a National Facility operated by the
University of Manchester at Jodrell Bank Observatory on behalf of
PPARC.  We acknowledge the continuous support by Phil Diamond, Harm
Habing and Richard Schilizzi.


\begin{chapthebibliography}{1}

\bibitem{colomer}
  Colomer F., Reid M.J., Menten K.M., Bujarrabal V., 2000, AA 355 979

\bibitem{diamond} 
  Diamond P.J., Kemball A.J., 1999, in ``Asymptotic Giant Branch Stars,
  IAU Symposium \#191'', eds Le Bertre, Lebre \& Waelkens, p.\ 195

\bibitem{eckart}
  Eckart A., Genzel R., Ott T., Sch\"odel R., 2002, MNRAS 331 917

\bibitem{marvel}
  Marvel, K.B., 1996, Ph.D. thesis, New Mexico State Univ

\bibitem{Mentgc}
  Menten K.M., Reid M.J., Eckart A., Genzel R., 1997, ApJ 475 L111

\bibitem{siva}
  Sivagnanam P., Diamond P.J., Le Squeren A.M., Biraud F., 1990, AA 229 171

\bibitem{sjouw}
  Sjouwerman L.O., van Langevelde H.J., Diamond H.J., 1998, AA 339 897

\bibitem{vlang92}
  Van~Langevelde, H.J., Frail, D.A., Cordes, J.M., Diamond, P.J., 1992, ApJ 396 686

\bibitem{vlang1}
  Van Langevelde H.J., Vlemmings W., Diamond P.J., Baudry A., Beasley
  A.J., 2000, AA 357, 945

\bibitem{vlemming}
  Vlemmings W.H.T., Van Langevelde H.J., Diamond P.J., 2002, AA {\sl in press}

\end{chapthebibliography}
\end{document}